%% file: trackmania.tex
\documentclass{article}
\usepackage{nips13submit_e,times}
\usepackage{hyperref}
\usepackage{url}
\usepackage{graphicx}
\usepackage{cite}
\usepackage{subfigure}
\usepackage{amsmath}
\usepackage{amssymb}
\usepackage{fancyheadings}
\usepackage{url}
\usepackage{hyperref}
\usepackage{algorithm}
\usepackage{algorithmic}
\usepackage{cancel}

\usepackage{float}
\floatstyle{ruled}
\newfloat{program}{thp}{lop}
\floatname{program}{Program}
\usepackage{placeins}

\usepackage{color}
\usepackage{xspace}

\usepackage{colortbl}

\usepackage{hyperref}
\hypersetup{
  colorlinks   = true,
  urlcolor     = blue,
  linkcolor    = blue,
  citecolor    = blue
}

\title{TrackMania is NP-complete}

\author{
Franck Dernoncourt \\
CSAIL, MIT\\
\texttt{francky@mit.edu}
}

\newcommand{\tmf}{TrackMania Nations Forever\xspace}

\nipsfinalcopy

\begin{document}

\maketitle

\vspace{0.4cm}

\textbf{Abstract}. We prove that completing an untimed, unbounded track in \tmf is NP-complete by using a reduction from 3-SAT and showing that a solution can be checked in polynomial time. 

\input{intro}

\bibliographystyle{unsrt}

\bibliography{trackmania}

\end{document}

%% file: intro.tex
\section{Introduction}

\tmf (TMNF, or TMF) is a 3D racing game that was released in 2008 by video game developer Nadeo. It is part of the racing game series TrackMania. It was designed for the Electronic Sports World Cup, which is a yearly international professional gaming championship that have distributed millions of dollars in prizes since its creation in 2003. Over 13 million online players signed up for TMNF, as the game is free of charge and its reception in video game magazines was largely positive. Guinness World Records \cite{guinness2008guinness}  awarded TrackMania six world records in 2008: ``biggest online race'', ``most popular online racing simulation'', ``most nationalities in an offline racing competition'', ``largest content base of any racing game'', ``first publicly available game developed specifically for an online competition'' and ``most popular user-created video''.

\begin{figure}[htb]
\centering
\includegraphics[width=315px]{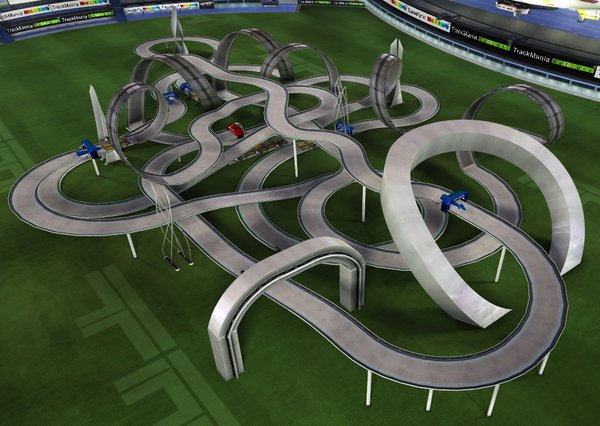}
\caption[3D track]{A track in \tmf (TMNF)}
\label{fig:track3d}
\end{figure}

In TMNF, the player's goal is to complete a track as quickly as possible. To complete a track, the player must first go through all checkpoints, which he can do in any order, then reach the finish gate. Figure \ref{fig:track3d} shows an example of a track.

This paper is, to the best of our knowledge, the first consideration of the
complexity of playing TrackMania or any other real-time racing game (see \cite{holzer2010computational} for a complexity proof for a non-real-time racing game).

\section{Rules of \tmf (TMNF)}

A track is composed of a set of 3D blocks that are positioned in a 3D space. The player's car appears on the start block. To complete the track, the player's car must go through all checkpoint blocks in any order, then reach the finish block. We ignore multi-lap tracks.

In addition to the start, checkpoint and finish blocks, there are typically many other road blocks, which can be straight roads, curves, slopes, dividers, etc. To enable the car to perform jumps between blocks or simply go faster, accelerator blocks can be placed, which as the name indicates increase the car's speed.

If the player is dissatisfied with his current time while racing through the track or is blocked somewhere, he can choose to respawn at the last checkpoint he went through or at the start block. Since the track is located in a 3D space the player's car may sometime fall and not be able to go back on the track, and might therefore need to respawn.

\begin{figure}[htb]
\centering
\includegraphics[width=400px]{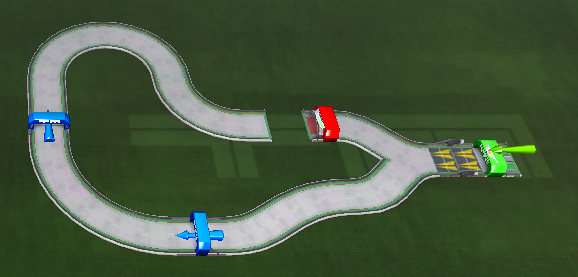}
\caption[Simple track example]{Example of a simple track. This track contains one start gate (green) at the right. The latter is adjacent to accelerator block (yellow). The player's car can either go to the finish block (red) or the two checkpoints (blue). In order to complete the track, the car must first go through the two checkpoints in any order, then go through the finish block.}
\label{fig:track01}
\end{figure}

We make two assumptions regarding the rules of TMNF:
\begin{itemize}
\item the track is \textit{untimed}, i.e. the player can take all the time he needs to complete the track.
\item the track is \textit{unbounded}, i.e. we ignore the fact that tracks have a maximum size.
\end{itemize}

Such relaxations are common in complexity proofs~\cite{kendall2008survey}.

\vspace{10pt}
\section{NP-completeness}

We present a reduction from 3-SAT, which is known to be NP-complete \cite{cook1971complexity}, to the problem of completing a track in TMNF. To that end, we present a variable gadget, a clause gadget and a crossover gadget. Any road can serve as wires.

\textbf{Variable gadget}

\begin{figure}[htb]
\centering
\includegraphics[width=400px]{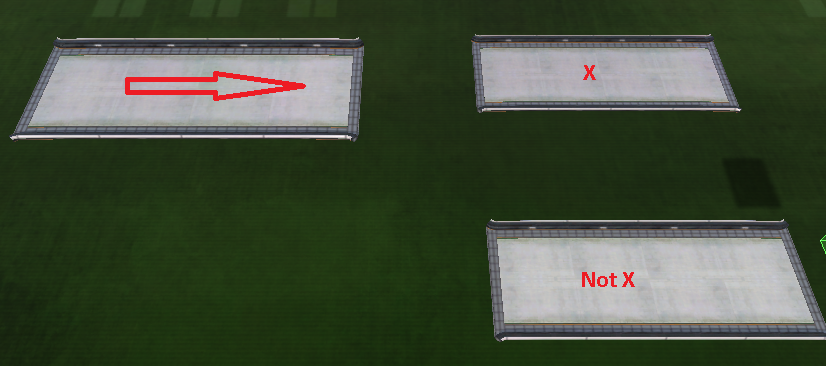}
\vspace{-16pt}
\caption[Variable gadget]{Variable gadget. It is composed of 3 platforms. The leftmost one is located in a higher altitude than the two rightmost platforms. The player's car must jump from the leftmost platform to one of the two rightmost platforms. The decision regarding which platform to jump to is similar to deciding whether a variable is true or false. Video: \url{http://youtu.be/-NO6ZWDqWYY}}
\label{fig:var-gadget-3d}
\end{figure}

\begin{figure}[htb]
\centering
\includegraphics[width=400px]{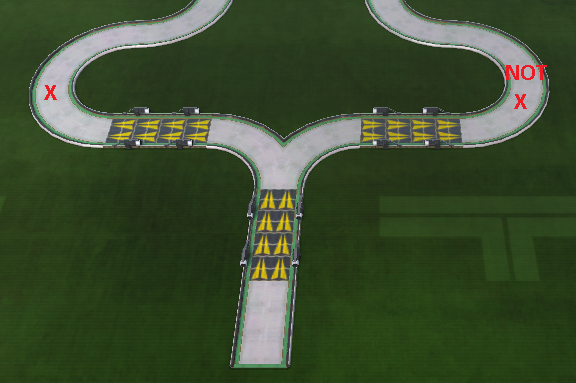}
\vspace{-16pt}
\caption[Variable gadget]{Variable gadget that does not work. It is tempting to make use of the  road accelerators to force the player to select a variable. However, a player's car can go through road accelerators in reverse direction, not matter how many subsequent road accelerators there are, as shown in the video \url{http://youtu.be/Em3am64t8LM}}
\label{fig:var-gadget-speed}
\end{figure}

Figure \ref{fig:var-gadget-3d} presents a variable gadget that uses the 3D property of the game. Figure \ref{fig:var-gadget-speed} does not work as a variable gadget due to a counter-intuitive ability of going through road accelerators in reverse direction.

\clearpage

\textbf{Clause gadget}

\begin{figure}[htb]
\centering
\includegraphics[width=400px]{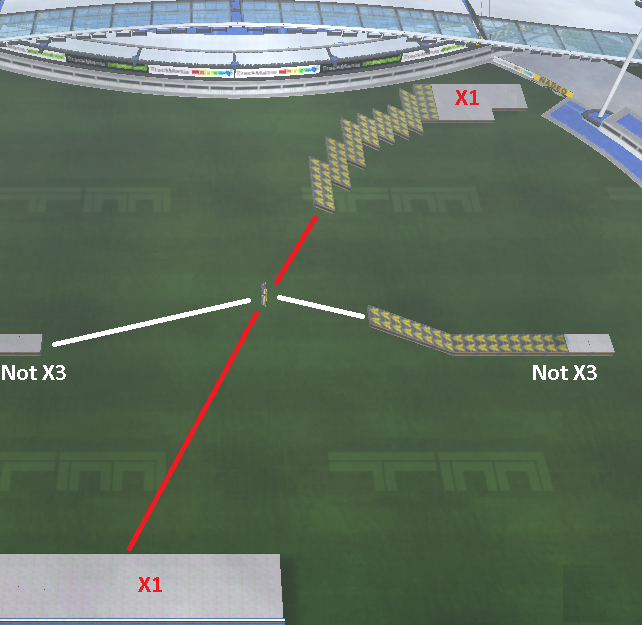}
\vspace{-16pt}
\caption[Clause gadget]{Clause gadget. The car comes from the right on one of the two paths. We only represented two paths in the figure due to space constraint, but in actuality the clause gadget contains three paths: the third path is axially symmetric to the top right path with respect to the middle right path. Each path represents a literal in the clause. Since we perform a reduction from 3-SAT, each clause has exactly 3 literals, which can be either positive or negative. In this example, we assume that the clause is $X_1 \wedge \neg X_3  \wedge X_4$ (2 positive literals and 1 negative literal). The $X_4$ literal is not represented on the figure. In this track's portion, the car comes from one of the three paths on the right, jumps through the checkpoint at the center in blue and lands on one of the three paths on the left. The path the driver lands on depends on the path he jumped from: the upper right path ($X_1$) can \textit{only} lead to the lower left path (\url{http://youtu.be/hj9PvWDNLvU}), the middle right path ($\neg X_3$) to the middle left path (\url{http://youtu.be/DnnXUJlOzDc}), and the lower right path ($X_4$) to the upper left path (not represented on the figure).
}
\label{fig:clause}
\end{figure}

The main idea behind the clause gadget is to place an ``aerial checkpoint'' so that it can be accessed through three different directions, to simulate the fact that if a literal is true then a clause that contains it is true: for each clause we create a checkpoint, which we insert in the track as shown in Figure \ref{fig:clause}.

\FloatBarrier

\clearpage

\textbf{Crossover gadget}

\begin{figure}[htb]
\centering
\includegraphics[width=300px]{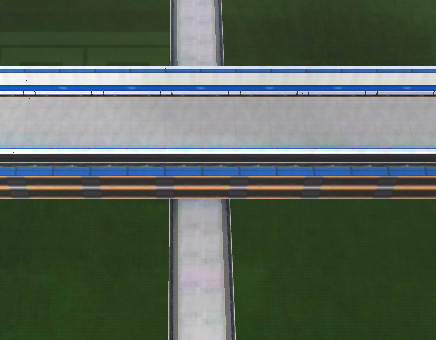}
\caption[Crossover gadget]{Crossover gadget. The player's car cannot fall from the higher path to the lower path due to the presence of barriers on each side of the road.}
\label{fig:crossover}
\end{figure}

Since we are in a 3D space, as a crossover gadget we can simply cross two paths at different altitude and make sure that the player cannot fall to the lower path, as shown in Figure \ref{fig:crossover}.

\vspace{15pt}

These 3 gadgets demonstrate that we can reduce from 3-SAT to TMNF. Furthermore all 3 gadgets can be constructed in polynomial time. This implies that TMNF is NP-hard. Since given a path in the track we can check in polynomial time whether it completes the track, it implies that TMNF is in NP. Since TMNF is NP-hard and in NP, it is therefore NP-complete.

\FloatBarrier

\vspace{0pt}

\section{Conclusion}
\vspace{-2pt}
We proved that the problem of completing a track in \tmf (TMNF) is NP-complete. To the best of our knowledge, the proof is also valid in other games of the TrackMania series (see Table~\ref{tab:tmseries}), as in all of them completing a track requires the car to go through all checkpoints in any order, then going through the finish block (put aside multi-lap tracks), and they contain the same or similar blocks\footnote{For example, the aerial checkpoint that we used was introduced in TrackMania United, and was not present before (e.g. TrackMania Nations): we can use a road checkpoint instead.} as the ones we used in the proof for TMNF. Proving NP-completeness leaves many questions open, such as refining the complexity analysis of tracks, or using the game to crowdsource tasks of higher (or lower, depending on one's point of view) interest\cite{walsh2014candy}.

\section{Acknowledgments}
\vspace{-2pt}
We thank Olivier Filipowicz, Ken L'H\'{e}ritier, and Jenny Lee for their fruitful ideas, as well as the Maniaplanet forum.

\section{Data}
\vspace{-2pt}
The tracks for each gadget can be found at \url{https://github.com/Franck-Dernoncourt/trackmania-np-complete}.

\section{The TrackMania series}

\begin{table}[H]
\centering
\begin{tabular}{ccc}
\hline 
Release date & Title & Platform\tabularnewline
\hline 
2003-11-28 & TrackMania & MS Windows\tabularnewline
2004-04-02 & TrackMania: Power Up! (Expansion Pack) & MS Windows\tabularnewline
2004-10-09 & TrackMania Speed-Up (Expansion Pack) & MS Windows\tabularnewline
2005-04-08 & TrackMania Sunrise & MS Windows\tabularnewline
2005-10-12 & TrackMania Original (Expansion Pack) & MS Windows\tabularnewline
2006-01-27 & TrackMania Nations ESWC & MS Windows\tabularnewline
2006-11-17 & TrackMania United & MS Windows\tabularnewline
2008-04-15 & Trackmania United Forever & MS Windows\tabularnewline
2008-04-16 & TrackMania Nations Forever  & MS Windows\tabularnewline
2009-03-17 & TrackMania DS & Nintendo DS\tabularnewline
2011-03-24 & TrackMania: Build To Race & Wii\tabularnewline
2011-04-19 & Trackmania Turbo & Nintendo DS\tabularnewline
2011-09-14 & TrackMania 2: Canyon & MS Windows\tabularnewline
2013-02-27 & TrackMania 2: Stadium & MS Windows\tabularnewline
2013-07-04 & TrackMania 2: Valley & MS Windows\tabularnewline
\hline 
\end{tabular}
\caption[The TrackMania series]{The TrackMania series. All games in the series comes with a track editor.}
\label{tab:tmseries}
\end{table}

\vspace{10pt}